\documentclass[fleqn,usenatbib]{mnras}

\usepackage{newtxtext,newtxmath}
\usepackage{amsmath}
\usepackage{natbib}
\usepackage{longtable, booktabs, threeparttablex}

\usepackage[T1]{fontenc}

\DeclareRobustCommand{\VAN}[3]{#2}
\let\VANthebibliography\thebibliography
\def\thebibliography{\DeclareRobustCommand{\VAN}[3]{##3}\VANthebibliography}

\usepackage{graphicx}	
\usepackage{amsmath}	

\title[Measuring the Mean Plane of the Distant Kuiper Belt]{Measuring the Mean Plane of the Distant Kuiper Belt}

\author[A. Siraj, C. F. Chyba, \& S. Tremaine]{
Amir Siraj,$^{1}$\thanks{E-mail: siraj@princeton.edu}
Christopher F. Chyba,$^{1,2}$
and Scott Tremaine$^{3}$
\\
$^{1}$Department of Astrophysical Sciences, Princeton University, 4 Ivy Lane, Princeton, NJ 08544, USA\\
$^{2}$School of Public and International Affairs, Princeton University, 20 Prospect Lane, Princeton, NJ 08540, USA\\
$^{3}$School of Natural Sciences, Institute for Advanced Study, Princeton, NJ 08540, USA\\
}

\date{Accepted XXX. Received YYY; in original form ZZZ}
 
\pubyear{\the\year{}}

\begin{document}
\label{firstpage}
\pagerange{\pageref{firstpage}--\pageref{lastpage}}
\maketitle

\begin{abstract}
In the absence of any unseen planetary-mass bodies in the outer solar system, the mean plane of the distant Kuiper belt should be the same as the plane orthogonal to the angular momentum vector of the solar system -- the invariable plane. Here, we measure the mean plane of the non-resonant Kuiper belt across semimajor axes $50 - 400 \mbox{\;AU}$. We introduce a new method to measure the mean plane that we demonstrate to be independent of observational bias. In particular, our results are not biased by surveys that look only at limited areas on the celestial sphere.
We find a warp relative to the invariable plane at semimajor axes of $80 - 400 \mbox{\;AU}$ ($98\%$ confidence) and $80 - 200 \mbox{\;AU}$ ($96\%$ confidence), but not at $50 - 80 \mbox{\;AU}$ or $200 - 400 \mbox{\;AU}$. If it is not spurious, a possible explanation for this warp is an unseen planet in the outer solar system. With $n$-body simulations, we find that a planet with mass between that of Mercury and the Earth, semimajor axis in the range $\sim 100 - 200 \mbox{\;AU}$, and inclination $\gtrsim 10^{\circ}$ to be the most likely cause of the warp; however, parameters outside of these ranges are still possible. Such a body is distinct in both mass and semimajor axis from the various versions of an unseen planet invoked to explain apsidal clustering in the outer solar system. The Vera C. Rubin Observatory's Legacy Survey of Space and Time (LSST) is expected to confirm or deny the existence of the warp reported here, and might detect the planet that may produce it.

\end{abstract}

\begin{keywords}
Kuiper belt: general -- minor planets, asteroids: general
\end{keywords}

\section{Introduction}

While there is a general consensus that the mean plane of the Kuiper belt is consistent with the invariable plane for $50\mbox{\;AU}\lesssim a \lesssim 80 \mbox{\;AU}$ \citep{2019AJ....158...49V, 2023AJ....165..241M}, not much is known about the mean plane of the Kuiper belt at $a \gtrsim 80 \mbox{\;AU}$. \cite{2023AJ....165..241M} found the mean plane of the Kuiper belt at semimajor axes of $50 - 150 \mbox{\;AU}$ to be consistent with the invariable plane; however, this bin is dominated in number by objects with semimajor axes in the range $50 - 80 \mbox{\;AU}$, so it is unclear what the contribution from objects with $a > 80 \mbox{\;AU}$ is. \cite{2020A&A...637A..87L} measured the mean plane of the Kuiper belt in the $100 - 200 \mbox{\;AU}$ semimajor axis range, finding it to be consistent with the invariable plane. However, they included objects in possible mean-motion resonances with Neptune, which may contaminate any signal of a warp. In addition, most previous work on measuring the Kuiper belt's mean plane uses the \cite{2004AJ....127.2418B} method -- such as the two aforementioned works -- or simple orbit-normal averaging. Both methods are subject to varying degrees of observational bias.

In this \textit{Letter}, we introduce a new method for measuring the mean plane of the Kuiper Belt and perform mean plane measurements at semimajor axes of $50 - 400 \mbox{\;AU}$. In Section \ref{meanplane}, we review the two existing methods for measuring the mean plane of a set of orbits, introduce our new method, and compare the performance of the three using simulated observations. In Section \ref{sec:results}, we measure the mean plane for non-resonant objects in the semimajor axis bins $50 - 80 \mbox{\;AU}$, $80 - 200 \mbox{\;AU}$, $200 - 400 \mbox{\;AU}$, as well as for the combined $80 - 400 \mbox{\;AU}$ bin. We confirm the findings of previous work regarding the $50 - 80 \mbox{\;AU}$ semimajor axis bin and report a significant deviation from the invariable plane in the $80-200 \mbox{\;AU}$ and $80-400 \mbox{\;AU}$ bins. In Section \ref{planet}, we investigate whether these results are consistent with unseen planets of various masses and orbits. Finally, in Section \ref{discussion}, we summarize our key findings and discuss implications of this work. Throughout this \textit{Letter}, we refer to small bodies in the solar system with $30 \mbox{\;AU} < a < 1000 \mbox{\;AU}$ and $q > 30 \mbox{\;AU}$ as trans-Neptunian objects (TNOs), but they are also commonly referred to as Kuiper belt objects (KBOs). All quoted distances are barycentric.

\section{Measuring the mean plane}
\label{meanplane}

The na\"ive approach to measure the mean plane of a sample of TNOs would simply be to take the average of all unit vectors normal to the plane of each TNO orbit. However, this method is highly subject to observational bias, as it does not differentiate between apparent clustering of the orbital planes due to on-sky selection effects and apparent clustering of the orbital planes due to the intrinsic distribution of orbits. Below, we describe two methods that mitigate the effects of observational bias on the mean plane measurement.

The \cite{2004AJ....127.2418B} approach, as implemented by \cite{2017AJ....154...62V} and \cite{2023AJ....165..241M}, mitigates the effects of observational bias by finding the plane that maximizes the symmetry of the sample's on-sky velocity vector distribution. 

Specifically, for each object in the sample, we calculate the unit vector in the direction of the on-sky velocity, 
\begin{equation}
    \mathbf{\hat v}_{t} = \frac{\mathbf{\hat v} - (\mathbf{\hat r} \cdot \mathbf{\hat v}) \mathbf{\hat v}}{|1-\mathbf{\hat r}\cdot\mathbf{\hat v}|} \; \; ,
\end{equation}
where $\mathbf{\hat v}$ and $\mathbf{\hat r}$ are the barycentric velocity and position unit vectors. We then find the unit vector $\mathbf{\hat n}$ that minimizes the quantity,
\begin{equation}
\label{minimize}
    \sum_{i = 1}^{N} (\mathbf{\hat n} \cdot \mathbf{\hat v}_{t, i})^2 \; \; ,
\end{equation}
where the sum is over all of the TNOs in a given semimajor axis bin. We require that the vector $\mathbf{\hat n}$ corresponds to an inclination angle in the range $0 - 90^{\circ}$ to avoid the degeneracy in which two vectors oriented in opposite directions describe the same plane. We minimize the sum of squares rather than the sum of absolute values (as was done by \citealt{2004AJ....127.2418B} and \citealt{2017AJ....154...62V}) so that there is a unique solution for each sample.\footnote{This distinction is illustrated in the following example: suppose we have two points with values $+1$ and $-1$. Then if we use the sum of squares, $\chi^2=(x-1)^2+(x+1)^2=2x^2+2$ which has a unique minimum at $x=0$. In contrast, $\chi^2=|x-1|+|x+1|$ is $2x$ for $x>1$, $-2x$ for $x<-1$, and 2 for $-1<x<1$. So the minimum of $\chi^2$ is not unique.} The vector $\mathbf{\hat n}$ is then the unit vector normal to the mean plane of the Kuiper belt in that semimajor axis bin. It is related to inclination and longitude of node as follows,
\begin{equation}
    \mathbf{\hat n} = (\sin{i_0}\sin{\Omega_0}, -\sin{i_0}\cos{\Omega_0}, \cos{i_0}) \; \; ,
\end{equation}
where $i_0$ and $\Omega_0$ are the inclination and longitude of node of the mean plane, respectively. The mean plane can alternatively be expressed by
\begin{equation}
    p_{0} = \sin(i_0 / 2) \sin{(\Omega_0)},\quad 
    q_{0} = \sin(i_0 / 2) \cos{(\Omega_0)}.
    \label{eq:pqdef}
\end{equation}

We note that it is difficult to estimate the uncertainty of the mean plane measurement produced by the \cite{2004AJ....127.2418B} method without relying on assumptions about the intrinsic orbital distribution of TNOs (see the procedures used in \citealt{2017AJ....154...62V} and \citealt{2023AJ....165..241M}).

We now present an alternative method that also allows a direct estimate of the mean‐plane uncertainty. Specifically, in steady state the TNO phase‐space probability distribution can be written as $F(E,\mathbf{J})$, where $E = \frac{1}{2}v^2 - \frac{GM_\odot}{r}$ is the energy and $\mathbf{J} = \mathbf{r}\times\mathbf{v}$ is the angular momentum. Assuming the disk’s symmetry axis is given by the unit vector $\hat{\mathbf{m}}$, the distribution can depend on the direction of $\mathbf{J}$ only through its angular distance from $\hat{\mathbf{m}}$. If this dependence is independent of $E$ and $J$ we can separate the distribution,
\begin{equation}
F(E,J,\hat{\mathbf{J}}\cdot\hat{\mathbf{m}})
= g(E,J)\,f(\hat{\mathbf{J}}\cdot\hat{\mathbf{m}}),
\end{equation}
where $g$ and $f$ are probability distributions. The function $g$ is spherically symmetric and $f$ incorporates the angular dependencies associated with the orientation of the disc.

Any anisotropic pointing or seasonal effects -- or other selection effects that depend on the latitude and longitude of the TNO -- are carried by the on-sky survey selection function $w(\mathbf r)$; we assume there is no residual selection on the direction of the proper‐motion vector at fixed rate. The joint probability of observing a TNO with coordinates $(\mathbf{r},\mathbf{v})$ is then 
\begin{equation}
w(\mathbf{r})\,g(E,J)\,f(\hat{\mathbf{J}}\cdot\hat{\mathbf{m}}).
\end{equation}
The conditional probability of measuring an angular‐momentum direction $\hat{\mathbf{J}}$ is then
\begin{equation}
C^{-1}\,f(\hat{\mathbf{J}}\cdot\hat{\mathbf{m}}),
\end{equation}
where
\begin{equation}
C = \int d\hat{\mathbf{J}} \;f(\hat{\mathbf{J}}\cdot\hat{\mathbf{m}}).
\end{equation}

For a sample $\{\hat{\mathbf{J}}_i\}_{i=1}^N$, the likelihood is

\begin{equation}
L \;=\; \prod_{i=1}^N
 \frac{w(\mathbf r_i)\,g(E_i,J_i)\,f(\hat{\mathbf J}_i\!\cdot\!\hat{\mathbf m})}
      {\displaystyle\int d\hat{\mathbf J}\;w(\mathbf r_i)\,g(E_i,J_i)\,f(\hat{\mathbf J}\!\cdot\!\hat{\mathbf m})}.
\end{equation}
which simplifies to,
\begin{equation}
\label{eq:likelihood}
L = \prod_{i=1}^N \frac{1}{C_i}\,f(\hat{\mathbf{J}}_i\cdot\hat{\mathbf{m}})
\end{equation} 
where $C_i$ is the per-object normalization constant. By conditioning on each object’s observed position $\mathbf r_i$ (and its energy and scalar angular momentum $E_i$ and $J_i$), both $w(\mathbf r_i)$ and $g(E_i,J_i)$ cancel in the likelihood, leaving only the intrinsic orbital‐plane term $f(\hat{\mathbf J}_i\!\cdot\!\hat{\mathbf m})$.

We adopt the von Mises distribution for the distribution of orbital planes because it is standard and algebraically convenient: 
\begin{equation}
f(\hat{\mathbf{J}}\cdot\hat{\mathbf{m}})
= \exp\bigl(\gamma\,\hat{\mathbf{J}}\cdot\hat{\mathbf{m}}\bigr);
\end{equation}
We experimented with other distributions and found the specific choice to be unimportant. Note that in the limit of large $\gamma$ the von Mises distribution gives us a razor-thin disk, and when $\gamma=0$ a spherical distribution.
The von Mises distribution leads to 
\begin{equation}
C = \int d\hat{\mathbf{J}} \;\exp\bigl(\gamma\,\hat{\mathbf{J}}\cdot\hat{\mathbf{m}}\bigr).
\end{equation}
 To evaluate the integral for $C$, we use spherical coordinates with $\hat{\mathbf{m}}$ along the positive $z$‐axis. The unit tangential velocity vector is 
\begin{equation}
\hat{\mathbf{v}}_t = \hat{\boldsymbol{\phi}}\cos\psi + \hat{\boldsymbol{\theta}}\sin\psi,
\end{equation}
and the unit angular momentum is
\begin{equation}
\hat{\mathbf{J}} = \hat{\mathbf{r}}\times\hat{\mathbf{v}}_t
= -\hat{\boldsymbol{\theta}}\cos\psi + \hat{\boldsymbol{\phi}}\sin\psi.
\end{equation}
Then $\hat{\mathbf{J}}\cdot\hat{\mathbf{m}} = \cos\psi\,\sin\theta$ and
\begin{equation}
\label{eq:C}
C = \int_0^{2\pi} d\psi \;\exp(\gamma\cos\psi\,\sin\theta)
= 2\pi\,I_0(\gamma\,\sin\theta),
\end{equation}
where $\theta$ is the angle between $\hat{\mathbf{r}}$ and $\hat{\mathbf{m}}$ and $I_0$ is a modified Bessel function. Using Eqs. \eqref{eq:likelihood} and \eqref{eq:C}, the log‐likelihood is just
\begin{equation}
\log L
= \gamma\,\hat{\mathbf{m}}\cdot\sum_{i=1}^N\hat{\mathbf{J}}_i
\;-\;\sum_{i=1}^N\log I_0(\gamma\,\sin\theta_i)
\;+\;\text{const.}
\end{equation}
Here $\cos\theta_i = \hat{\mathbf{m}}\cdot\hat{\mathbf{r}}_i$. We maximize $\log L$ subject to $|\hat{\mathbf{m}}|=1$ by varying $\hat{\mathbf{m}}$ and $\gamma$. Finally, the measurement’s significance relative to the invariable plane, expressed in terms of standard deviations, is $\sqrt{2\Delta \log L}$, where $\Delta \log L = \log L_{\rm max} - \log L_{\rm inv}$, where $L_{\rm max}$ is the maximum likelihood and $L_{\rm inv}$ is the likelihood at the location of the invariable plane ($i_{\rm inv} = 1.6^{\circ}$ and $\Omega_{\rm inv} = 107^{\circ}$ in ecliptic coordinates).

In order to verify that this method is truly bias-free (assuming that the distribution of planes follows the von Mises distribution)  and to compare the relative performance of this method to the \cite{2004AJ....127.2418B} and na\"ive orbit-normal-averaging methods, we employ a Monte Carlo simulation involving synthetic TNO populations and hypothetical survey footprints.

Each synthetic population consists of orbits with semimajor axes drawn randomly from a uniform distribution in the range $80 - 400 \mathrm{\; AU}$, perihelia from a uniform distribution in the range $30 - 60 \mathrm{\; AU}$, and argument of perihelion and mean anomaly from uniform distributions in the range $0 - 2\pi$. Orbital planes, which define inclination and longitude of node, are drawn from a von Mises distribution, $p(\mathbf{\hat J \cdot \hat n}) = \exp{(\mathbf{\gamma \hat J \cdot \hat n})}$ with $\gamma = 100$ about the invariable plane. Each hypothetical survey is composed of two $20^{\circ}\times20^{\circ}$ pointings for which the central ecliptic longitude ($\lambda$) of each pointing is drawn from a uniform distribution in the range $0 - 360^{\circ}$, and the central ecliptic latitude ($\beta$) of each pointing is drawn from a uniform distribution in $\sin(\beta)$ for $\mid \beta \mid \leq 20^{\circ}$. Finally, all hypothetical surveys are limited to $r < 100 \mathrm{\; AU}$ as an observability requirement.

\begin{figure}
 \centering
\includegraphics[width=\linewidth]{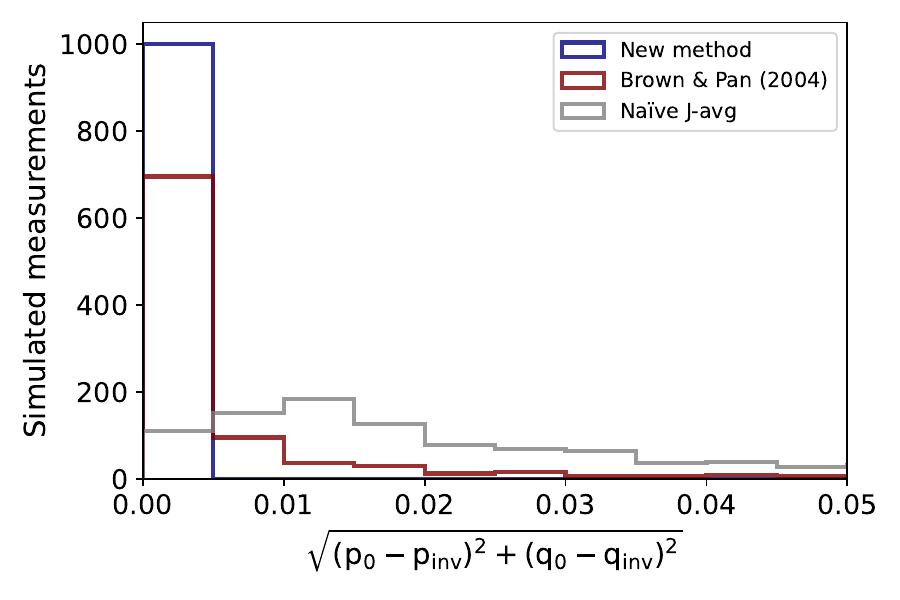}
\caption{Distribution of distances of simulated measurements of the mean plane from the actual mean plane for each of the three measurement methods in the Monte Carlo simulation described in Section \ref{meanplane}.}
\label{fig:hyp}
\end{figure}

We run $10^3$ instances of the Monte Carlo simulation in which we create a synthetic population with $10^7$ members, we `observe' said population with a randomly initialized hypothetical survey, and we `measure' the mean plane three ways, namely 1. our new, aforementioned method, 2. the \cite{2004AJ....127.2418B} method, and 3. na\"ive orbit-normal-averaging. The results are displayed in Figure \ref{fig:hyp}. We verify that the bias-free method works as expected and that the \cite{2004AJ....127.2418B} method is not perfect but works significantly better than simply taking the average of the orbit normal vectors. For instance, the distance to the true mean plane is less than a degree, 
$\sqrt{(p_0 - p_{\rm inv})^2 + (q_0 - q_{\rm inv})^2} < \sqrt{p(i = 1^{\circ})^2 + q(i = 1^{\circ})^2}$, for $100\%$ of the simulated measurements using the new method, $\sim 80\%$ using the \cite{2004AJ....127.2418B} method, and $\sim 20\%$ if one simply averaged the orbit normals.
Note that $p_{\rm inv} \equiv \sin(i_{\rm inv} / 2) \sin{(\Omega_{\rm inv})}$ and $q_{\rm inv} \equiv \sin(i_{\rm inv} / 2) \cos{(\Omega_{\rm inv})}$. We also ran simulations with lower values of $\gamma$ (for instance, $\gamma = 20$) and found consistent results.

\section{Results}
\label{sec:results}

We carry out this procedure for the $50 - 80 \mbox{\;AU}$, $80 - 200 \mbox{\;AU}$, $200 - 400 \mbox{\;AU}$, and $80 - 400 \mbox{\;AU}$ semimajor axis bins. Our sample was selected by first downloading all orbits in the JPL Small-Body Database (SBDB) as of May 14, 2025 with barycentric semimajor axis $50 \mbox{\;AU} < a < 400 \mbox{\;AU}$ and barycentric perihelion distance $q > 30 \mbox{\;AU}$. We exclude any TNOs with an excessively large uncertainty parameter\footnote{See \url{http://www.minorplanetcenter.org/iau/info/UValue.html}. The MPC `uncertainty parameter' is denoted the `condition code' in the JPL database.} ($> 6$) at the time of writing. This procedure gives us a sample of 740 TNOs. We then used \texttt{SBDynT}\footnote{\url{https://github.com/small-body-dynamics/SBDynT}} \citep{2024DPS....5620507V} to initialize and classify the orbits of 90 clones for each of the TNOs, with orbital elements of the clones drawn from probability distributions characterizing the observational uncertainties. If \texttt{SBDynT} classified any clone of a particular TNO as being in mean-motion resonance\footnote{The machine learning algorithm, the details of which are described in \cite{2024arXiv240505185V}, attempts to identify orbits which spend $\gtrsim50\%$ of the time undergoing libration in a resonant angle.} with Neptune, we exclude that TNO on the grounds of being possibly resonant, and therefore inappropriate to involve in a comparison to an expectation from secular theory -- namely, the null hypothesis that the measured mean plane of the distant Kuiper belt should be consistent with the invariable plane. This process resulted in a high-confidence sample of 154 non-resonant TNOs, which we fed into the procedures described in Section \ref{meanplane}. Note that if we relax the non-resonant criterion threshold (increase the allowed number of resonant clones) slightly, the significance of the warp described below increases, but if we relax it further, the significance decreases -- as might be expected if we are contaminating the sample with objects that are actually in resonance with Neptune.

\begin{table}
	\centering
	\caption{Non-resonant TNOs ($q > 30 \mathrm{\; AU}$) with semimajor axes in the range $80 \mbox{\;AU} - 400 \mbox{\;AU}$.}
	\label{tab:table_TNOs}
	\begin{tabular}{lcccr} 
		\hline
		TNO       & $a$ {[}AU{]} & $q$ {[}AU{]} & $i$ {[}deg{]} & $\Omega$ {[}deg{]}\\
		\hline
            2016 UP273 & 83.0  & 31.6 & 14.0     & 78.3       \\
            2013 GZ136 & 86.7  & 33.9 & 18.4     & 213.9      \\
            2008 JO41  & 87.4  & 39.9 & 48.8     & 153.1      \\
            2003 YQ179 & 88.5  & 37.3 & 20.9     & 109.8      \\
            1999 CF119 & 88.8  & 38.7 & 19.7     & 303.4      \\
            2015 TV361 & 88.8  & 37.9 & 25.2     & 33.6       \\
            2013 RK109 & 89.5  & 34.3 & 12.8     & 176.8      \\
            2013 RJ158 & 90.1  & 33.2 & 17.2     & 192.5      \\
            2012 HW87  & 91.7  & 32.5 & 14.1     & 125.5      \\
            2020 KV11  & 95.5  & 35.9 & 4.6      & 135.1      \\
            2011 BR163 & 97.3  & 37.9 & 24.8     & 107.4      \\
            2014 YD50  & 97.9  & 33.5 & 10.4     & 218.2      \\
            2010 ER65  & 98.1  & 40.0 & 21.3     & 212.6      \\
            2000 OM67  & 98.4  & 39.2 & 23.4     & 327.1      \\
            2013 VU71  & 100.6 & 34.2 & 13.9     & 141.0      \\
            2015 GY55  & 100.9 & 36.0 & 4.8      & 60.6       \\
            2015 RC279 & 101.3 & 31.7 & 18.1     & 358.1      \\
            1999 RZ215 & 101.5 & 31.0 & 25.5     & 341.6      \\
            2014 BE70  & 104.3 & 39.1 & 25.4     & 239.6      \\
            2014 VM43  & 107.1 & 32.9 & 16.7     & 157.4      \\
            2015 DW224 & 108.6 & 37.3 & 30.5     & 69.6       \\
            2017 BZ236 & 112.1 & 30.1 & 19.0     & 181.2      \\
            2019 SS149 & 116.8 & 35.3 & 14.9     & 94.5       \\
            2013 RE124 & 120.8 & 40.0 & 31.7     & 175.7      \\
            2014 MJ70  & 121.4 & 37.7 & 15.5     & 96.0       \\
            2014 KZ101 & 122.0 & 33.6 & 18.7     & 161.1      \\
            2015 RB279 & 126.4 & 33.5 & 19.8     & 192.4      \\
            2014 JW80  & 138.7 & 38.1 & 40.8     & 60.8       \\
            2013 JO64  & 143.3 & 35.1 & 8.6      & 185.2      \\
            2003 HB57  & 159.5 & 38.1 & 15.5     & 197.9      \\
            2015 SO20  & 164.8 & 33.2 & 23.4     & 33.6       \\
            2007 VJ305 & 192.0 & 35.2 & 12.0     & 24.4       \\
            2015 OC193 & 199.4 & 34.5 & 12.8     & 206.2      \\
            2017 CL54  & 210.7 & 36.0 & 14.9     & 201.6      \\
            2001 FP185 & 215.5 & 34.3 & 30.8     & 179.4      \\
            2015 RY245 & 225.1 & 31.3 & 6.0      & 341.5      \\
            2016 SG58  & 233.0 & 35.1 & 13.2     & 119.0      \\
            2023 KQ14  & 252.0 & 65.9 & 11.0     & 72.1       \\
            2012 VP113 & 262.3 & 80.5 & 24.1     & 90.8       \\
            2015 DY248 & 309.4 & 34.0 & 12.9     & 273.1      \\
            2015 GT50  & 311.2 & 38.4 & 8.8      & 46.1       \\
            2013 SL102 & 314.3 & 38.1 & 6.5      & 94.7       \\
            2014 TU115 & 335.3 & 35.0 & 23.5     & 192.3      \\
            2013 FL28  & 358.1 & 32.2 & 15.8     & 294.4      \\
            2013 RF98  & 358.9 & 36.1 & 29.5     & 67.6       \\
            2014 SX403 & 372.5 & 35.5 & 42.9     & 149.2      \\
		\hline
	\end{tabular}
\end{table}

\begin{figure*}
 \centering
\includegraphics[width=\linewidth]{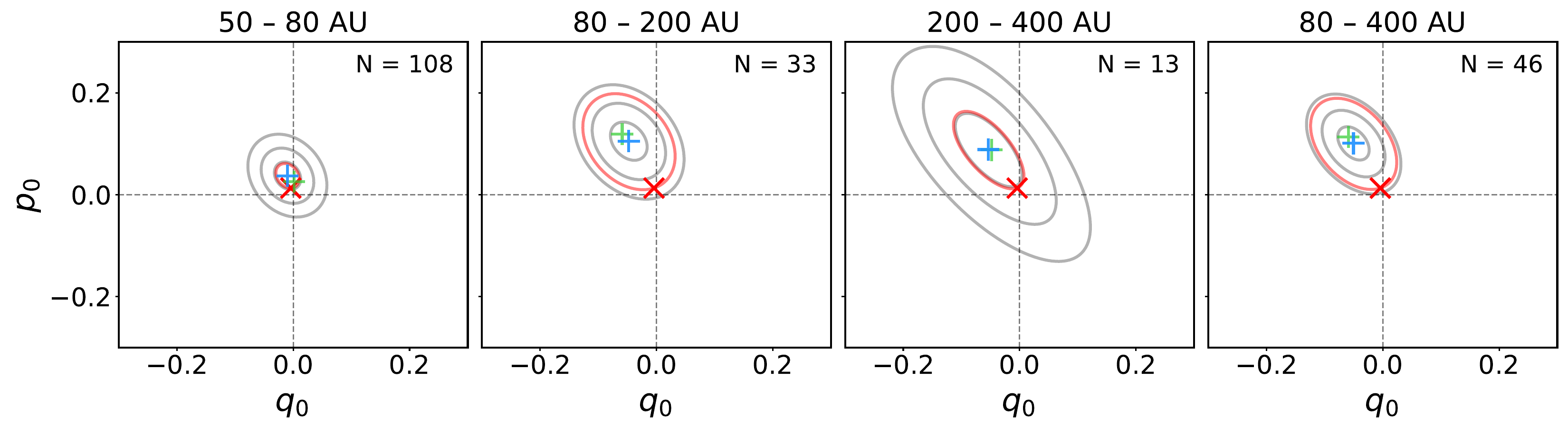}
\caption{Mean plane measurement using the new bias-free method described in Section \ref{meanplane} (blue pluses) and probability contour for the invariable plane (red ellipse) in the context of 1-, 2-, and 3-sigma probability contours (grey ellipses) for TNOs with perihelia $q > 30 \mbox{\;AU}$ and semimajor axes in the $50 - 80$ AU, $80 - 200$ AU, $200 - 400$, and $80 - 400$ AU bins. We have excluded objects for which any of the 90 clones sampled from the observational uncertainties of each object's orbit appeared to be in a mean-motion resonance with Neptune. The red `X' represents the invariable plane of the known planets. For comparison, measurements of the mean plane made with the \protect\cite{2004AJ....127.2418B} method are illustrated with green pluses. The fact that the invariable plane lies outside the area enclosed by the 2-sigma ellipse in the $80 - 200$ AU and $80 - 400$ AU bins illustrates the tentative detection of a local warp reported in this paper.}
\label{fig:T}
\end{figure*}

The results are illustrated in Figure \ref{fig:T} using the variables $p_0$ and $q_0$ defined in Equation (\ref{eq:pqdef}). The data are consistent with the invariable plane in the $50 - 80 \mathrm{\; AU}$ bin, as previous work has shown \citep{2019AJ....158...49V, 2023AJ....165..241M}, as well as in the $200 - 400 \mathrm{\; AU}$ bin. However, in the $80 - 200 \mbox{\;AU}$ or $80 - 400 \mbox{\;AU}$ bin the measurement of the mean plane is in $2.52 \sigma$ and $2.74 \sigma$ tension, respectively, with the invariable plane. Given the fact that our model has two free parameters, these levels of significance imply false alarm probabilities of $4\%$ and $2\%$, respectively. 

We have verified these false alarm probabilities by running Monte Carlo simulations of an un-warped disk and checking the fraction of cases in which the apparent warp exceeds the $\Delta \log{L}$ value of the observed warp. In these simulations, we distribute orbits symmetrically around the invariable plane, with the free components of the longitudes randomized from $0 - 360^{\circ}$ and those of $\sin{i}$ drawn from a Rayleigh distribution with dispersion $\sin{18^{\circ}}$. For each randomly initialized un-warped disk, we perform a mean plane measurement and record $\Delta \log{L}$. We find the overall fraction random initializations with $\Delta \log{L}$ exceeding the observed value to be $4\%$ and $2\%$ for the two bins, respectively, confirming the aforementioned false-alarm probability estimate. In contrast, for the $50 - 80 \mbox{\;AU}$ and $200 - 400 \mbox{\;AU}$ semimajor axis bins, invariable plane is only $\sim 1 \sigma$ away from the measured mean plane, and the false alarm probability is $\gtrsim 50\%$ -- indicating consistency between the mean plane measurements and the invariable plane. 

The warp has an approximate magnitude (inclination) of $i_0\sim 15^{\circ}$ and direction (longitude of node) of $\Omega_0\sim 120^{\circ}$, although the uncertainties on the magnitude and direction are quite large. The 46 non-resonant TNOs in the $80 - 400 \mbox{\;AU}$ bin are listed in Table \ref{tab:table_TNOs}. 

We confirm that small-number statistics and the particular positions of the 46 TNOs in question are unlikely to explain the observed warp, by creating a synthetic population (random $a$, $q$, and mean anomaly $M$ drawn from uniform distributions in the ranges of $80 - 400\mbox{\;AU}$, $30 - 50\mbox{\;AU}$, and $0 - 2\pi$, respectively) distributed symmetrically about the invariable plane with $\gamma = 20$ (representative of the best-fit $\gamma$ when both the mean plane and scatter are fit for the observed sample) and observing it with a hypothetical survey defined as the union of $10^{\circ}$ radius circles centered on each of the 46 TNOs. Running 100 trials comprising 46 `observed' objects each, the mean plane measurement described here yields $\sqrt{(p_0 - p_{\rm inv})^2 + (q_0 - q_{\rm inv})^2} < 0.07$ in all cases, making it unlikely for the observed warp in the $80 - 400\mbox{\;AU}$ bin (which is $0.1$ away from the invariable plane in $p-q$ space) to be a result of observational bias.

\section{An Unseen Planet?}
\label{planet}

The nodal precession rate is \citep{2019PhR...805....1B}

\begin{equation}
    \dot{\Omega} = -\frac{3}{4} \sqrt{\frac{GM_{\odot}}{a^3}} \frac{\cos{(i)}}{(1 - e^2)^2} \sum_{j = 5}^{8} \frac{m_j a_j^2}{M_{\odot} a^2} \; \; ,
    \label{eq:one}
\end{equation}
where $m_j$ and $a_j$ are the mass and semimajor axis of the $j$th planet from the Sun. For the TNOs in the $80 - 400 \mbox{\;AU}$ semimajor axis bin, the precession rate is $\sim 1 - 20 \mathrm{\; deg \; / \; Myr}$. This implies that any primordial warp in the mean plane would be washed away by differential precession in $\lesssim 10^8 \mbox{\;yr}$, meaning that any warp in the distant Kuiper belt requires some kind of ongoing gravitational shepherding in order to maintain the warp over the lifetime of the solar system. A possible explanation for the warp we report in the $80 - 200 \mbox{\;AU}$ and $80 - 400 \mbox{\;AU}$ semimajor axis bins -- if it is not spurious -- is that an unseen, inclined planet in or near this semimajor axis range is causing the mean plane of the Kuiper belt to deviate from the invariable plane. 

To explore the properties of an unseen planet that could induce a qualitatively similar warp, we ran a series of $1 \mathrm{\; Gyr}$ $n$-body integrations including the Sun, the four giant planets, the 154 non-resonant TNOs in the semimajor axis range $50 - 400 \mbox{\;AU}$ (treated as test particles), and an additional planet exterior to Neptune. At the start of each integration, we randomize the $\Omega$ values of the 154 non-resonant TNOs (drawing from a uniform distribution in the range $0 - 2\pi$) so that any warp observed later in the simulation must be attributable to forcing by the additional planet. We keep all other orbital elements the same as the present-day values for each TNO. For completeness, we also ran two control integrations in which no additional planet is included. In one, the $\Omega$ values of the 154 TNOs are left unaltered, and in the other, the $\Omega$ values are randomized across the interval $0 - 2 \pi$. One can make a simulated `measurement' of the mean planes in any semimajor axis bin at any timestep by  following the methodology outlined in Section \ref{meanplane}.

We carry out all $n$-body simulations using the \texttt{REBOUND} package \citep{2012A&A...537A.128R} with the \texttt{MERCURIUS} hybrid symplectic integrator \citep{2019MNRAS.485.5490R, 2024MNRAS.533.3708L}, using a timestep of 0.6 yr, which satisfies the \cite{2015AJ....150..127W} pericenter resolution criterion for all outer solar system orbits that do not interact strongly with Jupiter. This choice is suitable for the purposes of this work, since the focus is on objects with perihelion $q > 30 \mbox{\;AU}$, far larger than Jupiter's semimajor axis of $5.2\mbox{\;AU}$.

In the control integration in which the longitudes of node are unaltered, the orbits naturally isotropize in $\Omega$, as expected from the nodal regression induced by the giant planets (eq.\ \ref{eq:one}), and the mean planes across all four semimajor bins ($50 - 80 \mbox{\;AU}$, $80 - 200 \mbox{\;AU}$, $200 - 400 \mbox{\;AU}$, and $80 - 400 \mbox{\;AU}$) are consistent with the invariable plane once the warp in the $80 - 200 \mbox{\;AU}$ and $80 - 400 \mbox{\;AU}$ bins disappears. In the control integration in which the $\Omega$ values are randomized, the mean plane begins and remains consistent with the invariable plane during the entire integration in all four semimajor axis bins. We also ran separate control simulations including Pluto, Eris, and Sedna, verifying that those bodies cannot induce the observed warp in the $80 - 200 \mbox{\;AU}$ or $80 - 400 \mbox{\;AU}$ semimajor axis bin.

We now add an unseen planet. We first test nominal cases representing the planet favored by \cite{2021AJ....162..219B} ($m_p = 6.2 \mathrm{\; M_{\oplus}}$, $a_p = 380 \mbox{\;AU}$, $e_p = 0.2$, $i_p = 16^{\circ}$) and \cite{2025ApJ...978..139S} ($m_p = 4.4 \mathrm{\; M_{\oplus}}$, $a_p = 290 \mbox{\;AU}$, $e_p = 0.3$, $i_p = 7^{\circ}$). We find that neither planet can generate a warp of the required amplitude in the $80-200\mbox{\;AU}$ bin. However, we note that the current measurements of the mean plane of the Kuiper belt at semimajor axes of $80 - 400 \mbox{\;AU}$ are agnostic with respect to the existence of either planet. Significantly larger samples of objects would be necessary to refine the uncertainties on measurements of the distant Kuiper belt's mean plane enough to be able to accept or reject either hypothesis. Part of the problem is that the expected warp depends on the unseen planet's inclination, and this is quite uncertain in \cite{2021AJ....162..219B} and \cite{2025ApJ...978..139S}.

\begin{figure*}
 \centering
\includegraphics[width=\linewidth]{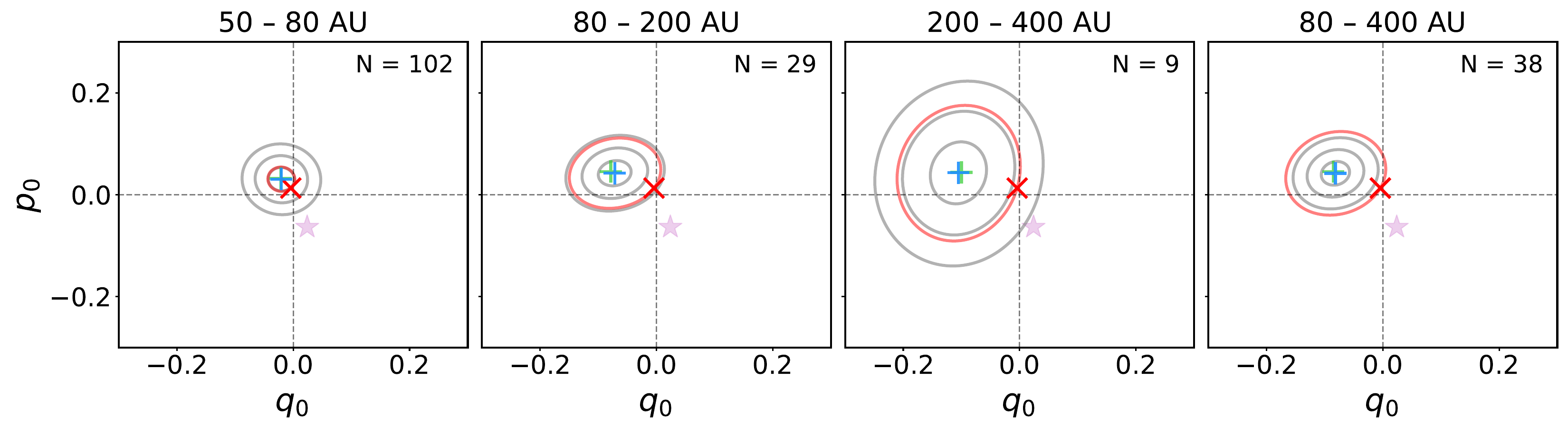}
\caption{Simulated mean plane measurement and probability ellipses for the same TNOs in Figure \ref{fig:T} after randomizing the $\Omega$ values and integrating for $500 \mathrm{\; Myr}$ with the giant planets and an additional planet with mass $m_p = 0.1 \mathrm{\; M_{\oplus}}$, semimajor axis $a_p = 130 \mbox{\;AU}$, perihelion $q_p = 120 \mbox{\;AU}$, and inclination $i_p = 10^{\circ}$. As in Figure \ref{fig:T}, the red `X' represents the invariable plane of the known planets, the blue plus marks the measurement of the mean plane using the new method described in Section \ref{meanplane}, and the green plus marks the measurement performed with the \protect\cite{2004AJ....127.2418B} method. The purple star in each panel represents the $(p_0, q_0)$ location of the orbital plane of the additional planet in the simulation. This is an illustrative example only.}
\label{fig:sim}
\end{figure*}

Given that a different unseen planet would be needed to explain the warp, we experiment with a wide variety of cases, namely masses in the range $10^{-3} - 4 \mathrm{\; M_{\oplus}}$, semimajor axes in the range $80 - 300 \mbox{\;AU}$, and inclinations in the range $5 - 55^{\circ}$. In our qualitative exploration, we ask whether an unseen planet can produce and maintain a warp in the mean plane of roughly similar amplitude and in the same semimajor axis bin as observed, while leaving the mean plane in the other semimajor axis bins consistent with the invariable plane. We find that planets with masses $> 1 \mathrm{\; M_{\oplus}}$ tend to significantly warp the mean plane of the $50 - 80 \mbox{\;AU}$ bin, making them poor candidates to match the observed warp. In the $\sim 0.1 - 1 \mathrm{\; M_{\oplus}}$ mass range, a subset of planets are able to produce a qualitatively similar warp in the $80 - 200 \mbox{\;AU}$ and $80 - 400 \mbox{\;AU}$ bins at $\sim 40\%$ of timesteps. In the $0.01-0.1\mathrm{\; M_{\oplus}}$ mass range, a subset of planets are able to similarly match observations $\sim 20\%$ of the time, and in the $0.001-0.01\mathrm{\; M_{\oplus}}$ mass range, a subset of planets are able to similarly match observations $\sim 10\%$ of the time. Therefore, while a planet with mass between that of Mercury and the Earth ($\sim 0.06 - 1\mathrm{\;M_\oplus}$) seems preferred to explain the warp, a Pluto-mass dwarf planet ($\sim 0.002\mathrm{\;M_\oplus}$) is not ruled out. In general, semimajor axes of $100 - 200 \mbox{\;AU}$ and inclinations of $\gtrsim 10^{\circ}$ tended to match the observations best. Figure \ref{fig:sim} shows an example of the `measured' mean plane across semimajor axis bins for an integration including an additional planet that shows similar qualitative behavior to the real measurements in Figure \ref{fig:T}.

Local warping of the mean plane is predicted by secular theory (see Appendix D of \citealt{2017AJ....154...62V} and Section 2 of \citealt{2020A&A...637A..87L}), but the actual effect in our $n$-body simulations appears to be much more complicated. In secular theory, the warp is given by the forced plane at each semimajor axis and the orientation of the forced plane is a superposition of terms from each planet. However, in our case the warp is forced almost entirely by the additional planet, for two reasons: (i) the inclinations of the known planets relative to the invariable plane are much smaller than the assumed inclination of the additional planet; (ii) both the additional planet and the TNOs we are examining have much larger semimajor axes than the known planets. 

In secular theory, the direction of the warp is expected to be aligned ($\Omega_p$) or anti-aligned ($\Omega_p + \pi$) with the planet -- due to the phase change induced across the secular resonance. In some cases the warp may be `washed out' if the semimajor axis distribution in a bin samples both the aligned and anti-aligned regimes. When a warp is observed in our $n-$body simulations (which does not occur $100\%$ of the time), the direction of the warp is not always aligned ($\Omega_p$) or anti-aligned ($\Omega_p + \pi$) with the planet, thus deviating from the secular theory expectation. However, the mass and semimajor axis constraints found here through $n$-body simulations are broadly consistent with expectations from linear secular theory.

\section{Discussion}
\label{discussion}

In this work, we introduced a new method for measuring the mean plane of the distant Kuiper belt in a bias-free way, and performed such a measurement on the population of non-resonant TNOs in the $50 - 400 \mbox{\;AU}$ semimajor axis range, confirming previous work that found no evidence for a warp in the $50 - 80 \mbox{\; AU}$ bin \citep{2019AJ....158...49V, 2023AJ....165..241M} and finding a moderately statistically significant local warp relative to the invariable plane in the $80 - 200 \mbox{\;AU}$ and $80 - 400 \mbox{\;AU}$ bins ($96\%$ and $98\%$ confidence, respectively). The warp reported here was not found in previous work reporting measurements of the mean plane of the Kuiper belt \citep{2017AJ....154...62V, 2020A&A...637A..87L, 2023AJ....165..241M, 2025arXiv250307146P}  partially due to one or more of the following issues: 1. not sufficiently excluding objects that may be in mean-motion resonance with Neptune, 2. not using a distant enough semimajor axis bin, 3. using simple orbit-normal averaging, which is subject to observational bias, or 4. using older, smaller and less accurate, catalogs of distant TNOs.

We are aware of the possibility that as the sample size in the $80 - 400 \mbox{\;AU}$ semimajor axis range continues to grow, the statistical significance of the observed warp could wane, as was the case for the warp reported by \cite{2017AJ....154...62V}, with similar significance to the warp found here, in the $50 - 80 \mbox{\;AU}$ bin \citep{2019AJ....158...49V, 2023AJ....165..241M}.

One possible explanation for a mean plane inconsistent with the invariable plane at semimajor axes of $80 - 400 \mbox{\;AU}$ is that the distribution of orbits in the distant Kuiper belt could reflect the gravitational influence of an unseen planet. We used $n-$body integrations to explore the plausibility of this scenario. In order to induce an effect similar to the one observed, the planet's orbit must be inclined with respect to the invariable plane. For the semimajor axis of the unseen planet, we find the range $100 - 200 \mbox{\;AU}$ to be most likely. A planet with a semimajor axis of $\lesssim 100 \mbox{\;AU}$ may produce significant forced inclinations in the $50 - 80 \mbox{\;AU}$ bin, and a planet with $a \gtrsim 200 \mbox{\;AU}$ may fail to produce a significant warp in the $80 - 200 \mbox{\;AU}$ bin. We find that a mass between that of Mercury and the Earth is favored ($\sim 0.06 - 1 \mathrm{\;M_\oplus}$), but a lunar- or Pluto-mass body ($\sim 10^{-3} - 10^{-2} \mathrm{\;M_\oplus}$) can also produce the observed warp, at least some of the time. Below $10^{-3} \mathrm{\; M_{\oplus}}$, it is difficult to excite a warp in the mean plane, and above $1 \mathrm{\; M_{\oplus}}$, the warp bleeds into the $50 - 80 \mbox{\;AU}$ bin. This upper limit is consistent with the finding that an Earth-mass body at such a distance from the Sun should have already been detectable in planetary ephemerides \citep{2023PSJ.....4...66G}.

Forced inclinations tend to be strongest near secular resonances, which occur somewhat outside the semimajor axis of the planet. However, we note that at semimajor axes near the semimajor axis of the planet, the forced inclination of TNOs is expected to be similar to the inclination of the planet, meaning that if the planet's inclination is large, this effect may potentially compete with that of the secular resonance. This becomes increasingly important for lower planet masses, since the width of the secular resonance is inversely related to planet mass.

The lower limit on semimajor axis is consistent with the stability requirement for TNOs in the 4:1 mean-motion resonance with Neptune, which may be disrupted in the case of relatively close-in unseen planetary mass bodies \citep{2023AJ....166..118L, 2025ApJ...978..139S}. A body like the one described here could have plausibly been produced in the early history of the solar system \citep{2006ApJ...643L.135G, 2018AJ....155...75S, Emel’yanenko2025}. An unseen planet motivated by the apsidal clustering of distant TNOs \citep{2014Natur.507..471T, 2021AJ....162..219B, 2025ApJ...978..139S} cannot explain, nor is ruled out by, the observational facts presented here. Such a planet is commonly referred to as Planet Nine or the historical term Planet X. We note that pre-emptive numbering (`Nine') is confusing when there is more than one independent hypothesis for unseen planets explaining different dynamical phenomena. Alternative terminology, such as Planet X to refer to a hypothesis explaining apsidal clustering (aforementioned work) and Planet Y to refer to a hypothesis explaining a warped mean plane (this work), may provide greater clarity. In such a framework, Planet X would refer to a distant ($a \gtrsim 200 \mbox{\;AU}$), high-mass ($M \gtrsim \mathrm{M_{\oplus}}$) planet while Planet Y would denote a closer-in ($a \lesssim 200 \mbox{\;AU}$), lower-mass ($M \lesssim \mathrm{M_{\oplus}}$) planet. While another option would be to refer to any hypothetical planet in the outer solar system as Planet X, multiple hypotheses for different hypothetical planets based on separate dynamical effects may be conflated if grouped under the same name.

Finally, we note that a hypothetical Planet Y as described in this work would likely be detectable by the upcoming Legacy Survey of Space and Time (LSST) on the Vera C. Rubin Observatory if it is currently located within the survey footprint \citep{2018AJ....155..243T, 2019ApJ...873..111I}. If such a body exists but is not discoverable by LSST due to its on-sky location (i.e., high ecliptic latitude), LSST will nevertheless elucidate the details of the Kuiper belt mean plane warp induced by the planet.

\section*{Acknowledgements}
We thank Renu Malhotra and Valentin Thoss for helpful comments that improved the quality of the manuscript. We are pleased to acknowledge that the work reported in this paper was substantially performed using the Princeton Research Computing resources at Princeton University which is consortium of groups led by the Princeton Institute for Computational Science and Engineering (PICSciE) and Office of Information Technology's Research Computing.

\section*{Data Availability}

The data underlying this article are available in the Jet Propulsion Laboratory Small Body Database (JPL SBDB), at \url{https://ssd.jpl.nasa.gov/tools/sbdb_query.html}.



\bibliographystyle{mnras}
\bibliography{bib} 

\bsp	
\label{lastpage}
\end{document}